# Numerical Investigation of Discontinuous Ice Effects on Swept Wings

Jiawei Chen[a], Maochao Xiao[b], Ziyu Zhou[a], YuFei Zhang[a,c,d]*

[a]School of Aerospace Engineering, Tsinghua University, Beijing 100084, China

[b]Department of Mechanical and Aerospace Engineering, Sapienza University of Rome, Rome 00184, Italy

[c]State Key Laboratory of Advanced Space Propulsion, Tsinghua University, 100084, Beijing

[d]Caofeidian Laboratory, 063200 Tangshan, China

* Correspondence: zhangyufei@tsinghua.edu.cn

**Abstract:** This study investigates the aerodynamic performance and flow structures of infinite swept wings with artificially simulated discontinuous ice using an enhanced delayed detached-eddy simulation. Comparisons are made among clean, continuous-ice, and discontinuous-ice configurations. Results show that discontinuous ice causes a more severe reduction in lift than continuous ice. While continuous ice forms a large separation bubble that helps maintain lift, discontinuous ice disrupts leading-edge vortex formation through gap jets, resulting in greater lift loss but a smaller drag penalty. Unlike the continuous-ice wing, the discontinuous-ice case does not exhibit a sudden stall-induced lift drop. The flow over the discontinuous-ice wing can be characterized by two canonical patterns: a separating shear layer and Kármán vortex shedding. However, the separating shear layer becomes irregular due to the interference of gap jets. Three characteristic chord-based Strouhal numbers ($St$)—11.3, 22.6, and 33.9—are identified. The lowest ($St = 11.3$) corresponds to the shedding of vortex pairs; when nondimensionalized by the ice width, it yields $St = 0.58$, which is higher than that of a canonical cylinder wake. Furthermore, lift and drag fluctuations occur predominantly at $St = 22.6$, twice the shedding frequency, primarily induced by the gap jets—a phenomenon absent in the continuous-ice case.

**Keywords:** ice accretion, delayed detached-eddy method, swept wing, discontinuous ice

## 1. Introduction

When an aircraft passes through clouds containing supercooled water droplets, ice can accumulate as the droplets freeze upon contact with the airframe. Ice accretion can severely degrade aerodynamic performance by reducing maximum lift, increasing drag, and diminishing control surface effectiveness[1]. For swept wings, the situation becomes more complex due to the formation of scallop ice[2]. Vargas[3] proposed and discussed various models and explanations for scallop formation. To quantitatively characterize the complex geometry of scallop ice, Wang et al.[4] proposed a geometric model based on experimental scallop ice. Understanding the formation mechanism and flow characteristics associated with scallop ice is essential for advancing the study of ice accretion and its aerodynamic consequences.

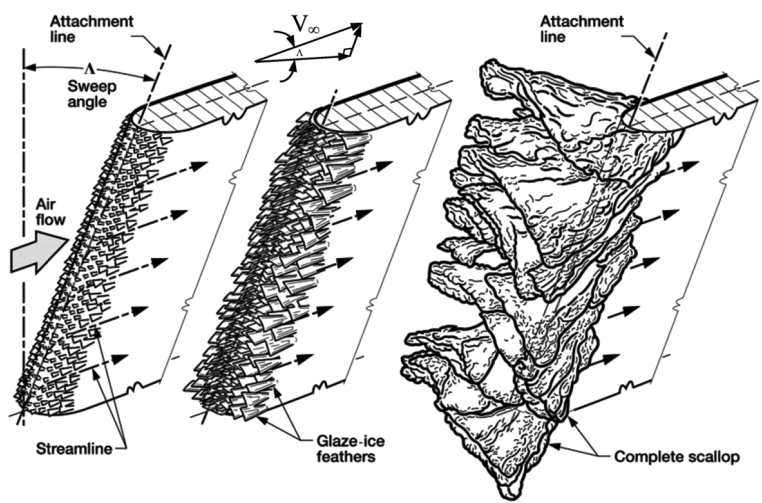

**Fig. 1. Ice accretion on a swept wing at glaze-ice conditions, complete scallop[3]**

Most existing studies have focused on simplified or span-averaged ice geometries. Bragg et al.[5][6][7] conducted extensive experimental studies and showed that the separation bubble downstream of glaze ice is significantly larger than that of rime ice[8], primarily due to the geometry of the ice horn. Such large-scale separation poses a major challenge for conventional linear eddy-viscosity models. Li et al.[9] modified the $k - \overline{v^2} - \omega$ model to improve predictions for rime and glaze ice. Chen et al.[10] enhanced the $\gamma - Re_{\theta t}$ transition model by introducing a separating shear-layer correction, extending its applicability to roughness-induced transition, and further employed it in ice-growth simulations[11]. Nevertheless, RANS-based approaches fail to capture the inherently unsteady dynamics of separated flows[12][13]. In contrast, hybrid RANS/large-eddy simulations (LES) methods and LES are more suitable for resolving detailed flow structures. Xiao et al.[14] applied the improved delayed detached-eddy simulation to study the flow field around horn and ridge ice and examined the influence of different subgrid length scales on the prediction. Wong et al.[15] conducted wall-modeled large-eddy simulations of the NACA 23012 airfoil with horn ice accretion, revealing that Kelvin–Helmholtz instability triggered by the upper ice horn leads to rapid laminar-to-turbulent transition and significantly affects aerodynamic performance across a wide range of angles of attack. Lee et al.[16] conducted LES of multi-element iced airfoils under supercooled large droplet (SLD) and non-SLD conditions, demonstrating that flow interactions near the slat gap play a crucial role in determining aerodynamic degradation and that LES provides significantly more accurate solutions than URANS for complex flows around iced airfoils.

Recently, increasing attention has been devoted to flow fields associated with realistic, high-fidelity ice accretion. Because generating grids for high-fidelity ice geometries is challenging, most existing studies have relied primarily on experimental approaches. Diebold et al.[17] investigated the aerodynamic impact of high-fidelity ice on swept wings at low Mach numbers and observed that flow separation at the ice tip can trigger the formation of a leading-edge vortex at low angles of attack. Sandhu et al.[18] examined how the fidelity of simulated ice shapes influences both the flow field and aerodynamic performance, showing that gaps in the ice produce streamwise jets that interfere with the formation of leading-edge vortices. Woodard et al.[19] used artificially designed discontinuous ice shapes on a swept wing to mimic the effects of realistic accretions. Reghin et al.[20] investigated the aerodynamic effects of artificial scallop ice shapes on a NACA 23012 airfoil, demonstrating that increasing gap widths in three-dimensional ice formations directly improves airfoil performance and that PIV flow-field analysis reveals flow reattachment downstream of horn bubbles for ice shapes with gaps. In contrast to the extensive body of experimental investigations, numerical studies remain relatively scarce. Chen et al.[21] employed a modified $\gamma - Re_{\theta t}$ transition model to investigate aerodynamic and flow-pattern differences between continuous and discontinuous ice on straight and swept wings. Bornhoft et al.[22] employed wall-modeled LES to simulate flow over a NACA 23012 airfoil with detailed ice structures, highlighting the necessity of accurately resolving ice geometry and surface roughness in numerical simulations.

This study employs the IDDES method with an enhanced subgrid-scale length formulation based on the anisotropic minimum-dissipation (AMD) model[23] to investigate the flow characteristics around discontinuous ice. Zhou et al.[24] demonstrated that the AMD-based formulation effectively handles anisotropic grids and substantially mitigates the "gray area" problem; it has also been successfully applied to iced-wing simulations[25]. Building on this foundation, the present work further explores the flow mechanisms of discontinuous ice. An infinite-span wing is adopted to eliminate wing-root and wing-tip influences. The study examines aerodynamic coefficients, flow structures, and characteristic frequencies in detail. Particular attention is given to the vortex-shedding frequency induced by discontinuous ice and its correlation with fluctuations of aerodynamic force coefficients.

The remainder of this paper is organized as follows. Section 2 introduces the computational methodology, including the IDDES with anisotropic minimum-dissipation subgrid length, numerical methods, and validation of the AMD-IDDES method. Section 3 presents the results for the infinite-span iced swept wing. Finally, Section 4 provides the conclusions of this study.

## 2. Computational methodology

### A. IDDES with anisotropic minimum-dissipation subgrid length

The AMD-IDDES method[25] is an enhanced variant of the standard IDDES approach, developed to mitigate the "gray area" problem and improve the simulation of complex flows, particularly those involving separation and transition. It integrates the IDDES framework with the AMD model[23], which dynamically adjusts the subgrid length scale based on local flow features. Originally proposed by Rozema et al.[23], the AMD model provides the minimum necessary eddy dissipation to remove subfilter-scale energy, locally approximating the exact dissipation and aligning with the nonlinear gradient model. By incorporating these refinements, the AMD-IDDES method significantly improves the prediction of critical flow phenomena such as separation bubbles, vortex shedding, and reattachment, especially in strongly anisotropic regions.

The IDDES method, based on the two-equation shear stress transport (SST) $k - \omega$

model, adjusts the transport equation for turbulent kinetic energy by replacing the RANS length scale in the destruction term with the IDDES length scale[26][27]. Specifically, the transport equation for $k$ in IDDES is given by:

$$\frac{\partial(\rho k)}{\partial t}+\frac{\partial(\rho u_j k)}{\partial x_j}=P_k-\frac{\rho k^{\frac{3}{2}}}{l_{IDDES}}+\frac{\partial}{\partial x_j}\left[(\sigma_k \mu_t+\mu)\frac{\partial k}{\partial x_j}\right] \quad (1)$$

where $\rho$ is the density, $u$ is the velocity, $k$ is the turbulent kinetic energy, and $\sigma_k$ is the turbulent diffusion constant. $\mu$ and $\mu_t$ are the molecular viscosity and turbulent viscosity, respectively. The production term is given by $P_k = \mu_t S$, where $S = \sqrt{2S_{ij}S_{ij}}$ is the invariant of the strain rate tensor.

The IDDES length scale is defined as

$$l_{IDDES}=f_d(1+f_e)l_{RANS}+(1-f_d)l_{LES} \quad (2)$$

where $l_{RANS}=\sqrt{k}/(C_\mu \omega)$ and $l_{LES}$ represent the RANS scale and LES length scale, respectively. The blending function $f_d$ ranges between 0.0 (LES mode) and 1.0 (RANS mode). The LES subgrid scale is given by:

$$l_{LES}=\min\{l_{wall},l_{free}\} \quad (3)$$

where $l_{wall}=C_w \max[d_w,\Delta_{\max}]$ and $l_{free}$ are the LES subgrid length scales in the near-wall vicinity and the wall-free regions, respectively. Here, $C_w$ is a constant, $d_w$ is the distance to the nearest wall, and $\Delta_{\max}$ is the maximum length of a cell edges.

In AMD-IDDES, $l_{free}$ is taken as the AMD subgrid length scale. The eddy viscosity of the AMD model is expressed as:

$$\nu_t = C_A^2 \frac{-\Delta_k g_{ik} \Delta_k g_{jk} S_{ij}}{g_{ml} g_{ml}} \tag{4}$$

where $C_A$ is a constant, $g_{ij} = \partial u_i/\partial x_j$, and $S_{ij} = (g_{ij} + g_{ji})/2$. Mathematically, $\Delta_k g_{ik}$ represents the scaled velocity gradient. When assuming an equilibrium state for subgrid turbulence, the eddy viscosity of the LES branch in the IDDES method, based on the SST $k-\omega$ turbulence model, is expressed as:

$$\nu_t = \left(\frac{\gamma}{\beta}\right)^{\frac{3}{2}} l_{LES}^2 S \tag{5}$$

where $\gamma$ and $\beta$ are the parameters in the production term and destruction term in the SST $k-\omega$ model. $S = \sqrt{2 S_{ij} S_{ij}}$ is the invariant of strain rate tensor $S_{ij}$. The AMD eddy viscosity can be reformulated using the IDDES approach as follows:

$$l_{free} = C_A \left(\frac{\gamma}{\beta}\right)^{\frac{3}{4}} \left(\frac{-\Delta_k g_{ik} \Delta_k g_{jk} S_{ij}}{g_{ml} g_{ml} S}\right)^{\frac{1}{2}} \tag{6}$$

Or equivalently

$$l_{free} = C_{DES,AMD} \Delta_{AMD} \tag{7}$$

$$C_{DES,AMD} = C_A \left(\frac{\gamma}{\beta}\right)^{\frac{3}{4}} \tag{8}$$

$$\Delta_{AMD} = \left(\frac{-\Delta_k g_{ik} \Delta_k g_{jk} S_{ij}}{g_{ml} g_{ml}}\right)^{\frac{1}{2}} \tag{9}$$

Rozema suggested that $C_A^2 = 3.0$ provides a suitable choice for a second-order central-difference scheme. This value yields $C_{DES,AMD} = 1.92$ with $\gamma = 0.44$ and $\beta = 0.0828$. In our simulations, this parameter is calibrated as $C_{DES,AMD} = 2.40$ using decaying isotropic turbulence following Zhou et al.[24].

Furthermore, a lower limit $C_{lim}V^{1/3}$ is imposed on the length scale to ensure a smaller LES length scale when the "gray area" issue arises. $C_{lim} = 0.15$ is an empirical constant, $V = \Delta_x \Delta_y \Delta_z$ is the volume of a grid cell. In RANS-modeled attached boundary layers, the LES length scale can become so small that the hybrid length scale falls below the RANS length scale. Consequently, RANS functionality could be compromised, yielding lower wall-friction values. To circumvent these potential defects, the LES length scale is restricted as:

$$l_{free} = \left(C_{DES,AMD}\Delta_{AMD}\right)_{lim} = \begin{cases} \max\left(C_{DES,AMD}\Delta_{AMD}, C_{lim}V^{1/3}\right), & f_d \leq \varepsilon \\ C_{DES}\Delta_{max,} & f_d \geq \varepsilon \end{cases} \quad (10)$$

where $C_{DES}$ is the parameter from the standard IDDES method, and the threshold value $\varepsilon$ is set to 0.01. The ultimate LES length scale becomes:

$$l_{LES} = \min\left(C_w \max[d_w, \Delta_{max}], \left(C_{DES,AMD}\Delta_{AMD}\right)_{lim}\right) \quad (11)$$

Unlike the AMD-IDDES method, the standard IDDES method employs the maximum cell spacing scale for the subgrid scale, where $l_{free} = C_{DES}\Delta_{max} = C_{DES}max(\Delta_x, \Delta_y, \Delta_z)$, with $\Delta_x, \Delta_y$ and $\Delta_z$ being the grid scales along the three coordinate directions. In real engineering problems, cell dimensions in the three directions can vary significantly. Figure 2 illustrates two types of anisotropic grids: the pencil ($\Delta_x \sim \Delta_y \ll \Delta_z$) cell and the book cell ($\Delta_y \ll \Delta_x \sim \Delta_z$). The standard IDDES method can be sensitive to highly anisotropic cells, whereas the AMD-IDDES method is more robust and applicable to practical engineering cases.

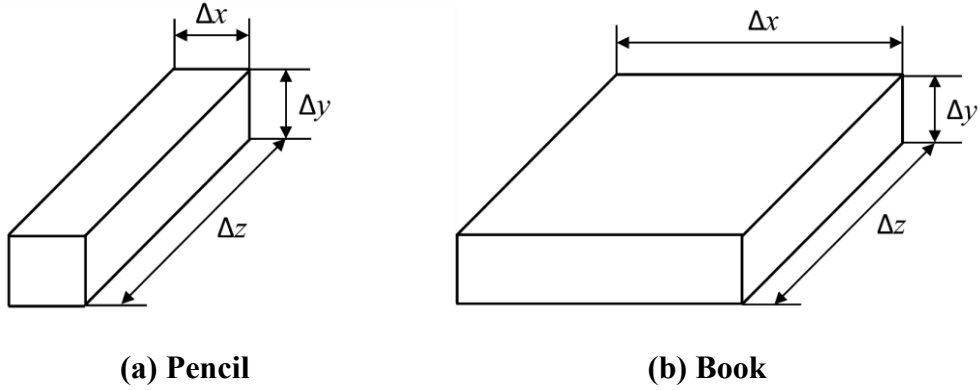

(a) Pencil  (b) Book

Fig. 2. Two types of anisotropic grids in engineering applications

## B. Numerical methods

The Navier–Stokes equations are solved with CFL3D[28], a structured solver based on the finite-volume method. Time integration employs a dual-time-stepping approximate factorization scheme[29], incorporating sub-iterations and multigrid techniques to achieve second-order accuracy and accelerate pseudo-time convergence. The viscous flux is discretized using a second-order central-difference scheme, while the inviscid flux is computed with a hybrid central/upwind scheme.

$$F_{invisid} = (1-\sigma) F_{central} + \sigma F_{upwind} \qquad (12)$$

where the central flux $F_{central}$ is discretized using a fourth-order central difference scheme, while the upwind flux $F_{upwind}$ is computed with the Roe scheme in conjunction with the third-order MUSCL scheme[30]. The blending function is given by:

$$\sigma = \max\left( \tanh\left( \frac{C_3}{1-C_4} \max\left( \frac{l_{IDDES}}{l_{RANS}} - C_4, 0 \right) \right), \sigma_{\min} \right) \quad (13)$$

where $C_3 = 4.0$ and $C_4 = 0.6$ [31]. In separated regions, the ratio $l_{IDDES}/l_{RANS}$ is generally below 0.6[14], resulting in $\sigma = \sigma_{min}$. In this study, $\sigma_{min}$ is specified as 0.05.

## C. Validation of AMD-IDDES method

A NACA 0012 semispan wing with simulated glaze ice is selected to validate the AMD-IDDES method against the baseline IDDES method. The wing has a chord length of $c$ and an aspect ratio of $b/c = 2.0$. Icing tests were conducted under a freestream velocity of 58.12 m/s, an angle of attack of 4°, an icing duration of 5 min, a droplet median volume diameter of 20 μm, a liquid water content of 2.1 g/m³, and an ambient temperature of 18 °F[32][33]. The resulting ice accretion profile is defined along the chordwise direction, and flow data are extracted from five spanwise sections at $z/b$=0.27, 0.42, 0.56, 0.72, and 0.89.

Figure 3 shows the computational domain and boundary conditions for the iced NACA 0012 swept wing. The computational domain, consistent with that used by Li et al.[9], extends $7.5c$ in the streamwise, $2.8c$ in the spanwise, and $2.1c$ in the vertical directions. The Reynolds number based on the chord length and inflow velocity is $Re = 1.5 \times 10^6$, the Mach number is $Ma = 0.2$, and the angle of attack is 8°. At the domain boundaries, inflow and outflow conditions are imposed at the streamwise ends, while sidewalls are adiabatic and no-slip, except for a slip wall positioned upstream of the model leading edge. The nondimensional time step is $\Delta t U_\infty/c = 0.001$. Each simulation is advanced for 30 convective time units (CTUs), where $CTU = c/U_\infty$, and the final 25 CTUs are used for statistical averaging.

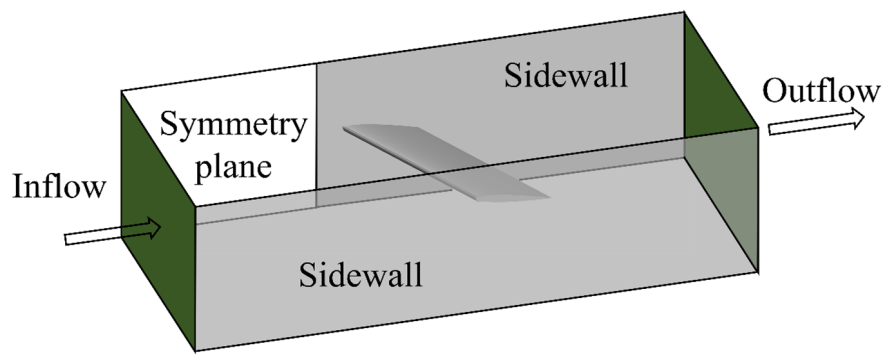

Fig. 3. Computational domain and boundary conditions for the iced NACA0012 swept wing

Table 1 describes the two sets of grids used in the study. The total grid numbers $N_{total}$ for the coarse and fine grids are $59 \times 10^6$ and $120 \times 10^6$, respectively. $N_x$, $N_y$, and $N_z$ denote the grid numbers in the streamwise, normal, and spanwise directions, respectively. The streamwise grid spacing in the ice accretion region is denoted by $\Delta x_{ice}/c$, while $\Delta y_w/c$, $\Delta z_w/c$ indicate the minimum wall-normal and spanwise grid spacings, respectively. Figure 4 shows the fine grid for the iced NACA0012 swept wing.

**Table 1: Details of two grid sets**

| Grid | $N_x$ | $N_y$ | $N_z$ | $\Delta x_{ice}/c$ | $\Delta y_w/c$ | $\Delta z_w/c$ | $N_{total}$ |
|---|---|---|---|---|---|---|---|
| Coarse | 532 | 172 | 476 | 0.0020 | $5 \times 10^{-5}$ | $5 \times 10^{-5}$ | $59 \times 10^6$ |
| Fine | 756 | 224 | 552 | 0.0013 | $5 \times 10^{-5}$ | $5 \times 10^{-5}$ | $120 \times 10^6$ |

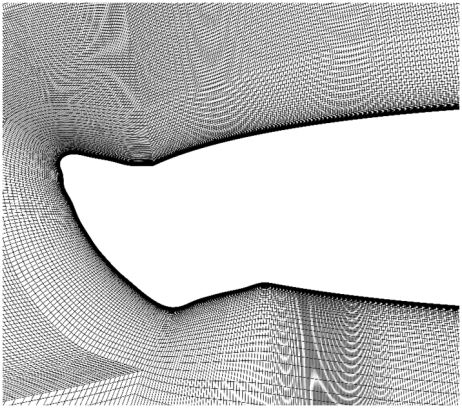 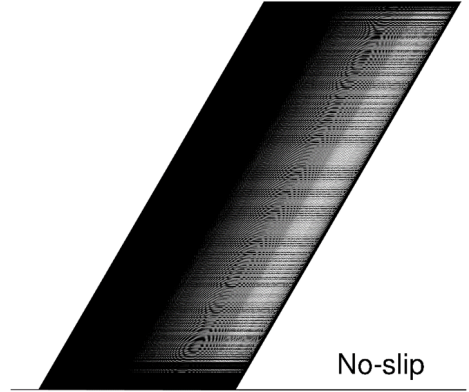

**(a) Leading-edge ice grids**  **(b) Spanwise grids**

**Fig. 4. Fine grids for the iced NACA0012 swept wing**

Figure 5 shows lift coefficients for the iced NACA0012 wing obtained from experiment[32][33] and AMD-IDDES computation. The AMD-IDDES method shows good agreement with the experimental data, accurately capturing the spanwise variation of the lift coefficient. Figure 6 shows time-averaged pressure coefficients for the iced swept wing obtained using the coarse and fine grids. Comparisons are made at three spanwise sections (*z*/*b* = 0.27, 0.56, and 0.89). The AMD-IDDES predicts a high-suction plateau, showing good agreement with the experimental data. For both grid resolutions, the AMD-IDDES demonstrates good accuracy, although some differences between the two grids can still be observed.

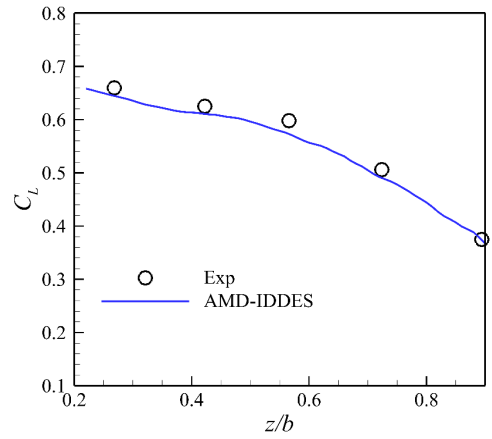

**Fig. 5. Lift coefficients for the iced NACA0012 wing obtained from experiment[32][33] and AMD-IDDES computation**

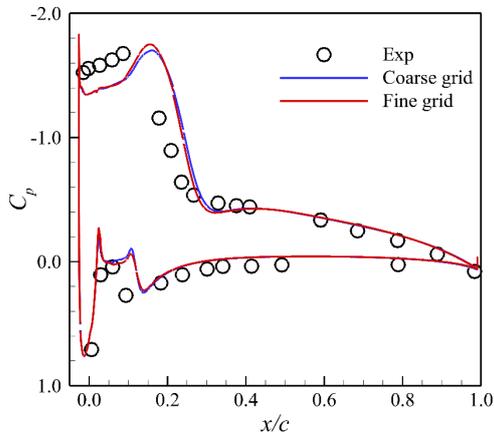

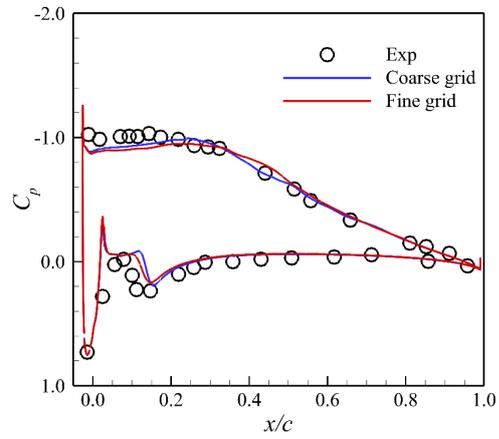

      (a) $z/b = 0.27$                       (b) $z/b = 0.56$

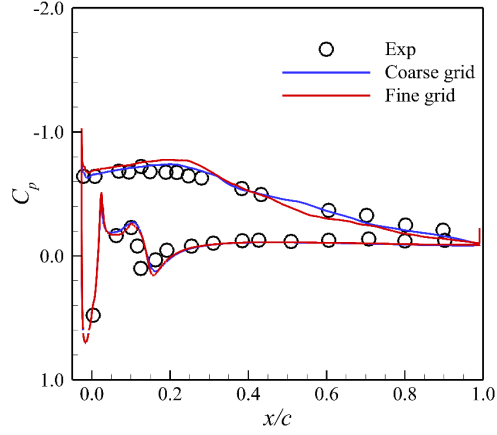

(c) $z/b = 0.89$

Fig. 6. Time-averaged pressure coefficients for the iced swept wing obtained using coarse and fine grids

## 3. Results and discussion

A NACA 23012 airfoil with an artificial discontinuous ice shape[34], as used by Reghin et al.[20], is adopted in this study. To focus on the flow characteristics induced by discontinuous ice accretion on swept wings, the finite-span wing geometry in Reghin et al.[20] is modified into an infinite-span swept wing configuration, thereby eliminating the influence of wing-root and wing-tip effects. The NACA23012 swept wing has a chord length $c$, an aspect ratio of $b/c = 0.93$, and a sweep angle of 30°. The ice shape has a relative height of $h/c = 0.0667$ and an incline angle of 54°, positioned on the upper surface of the airfoil at a relative distance of $h/c = 0.034$ from the leading edge. The Reynolds number based on the chord length and inflow velocity is $Re = 2.2 \times 10^5$. To validate the applicability of the AMD-IDDES method at this Reynolds number and to determine an appropriate grid resolution for the following infinite-span configurations, additional simulations are performed for the original finite-span wing geometry of Reghin et al.[20]. Figure 7 shows aerodynamic coefficients predicted using two different grid resolutions. Two meshes, referred to as the coarse and fine grids, are employed for comparison. The coarse grid has $25 \times 10^6$ cells, while the fine grid consists of $69 \times 10^6$ cells. As the grid resolution increases, the predictions obtained from the fine grid show improved accuracy compared with those from the coarse grid.

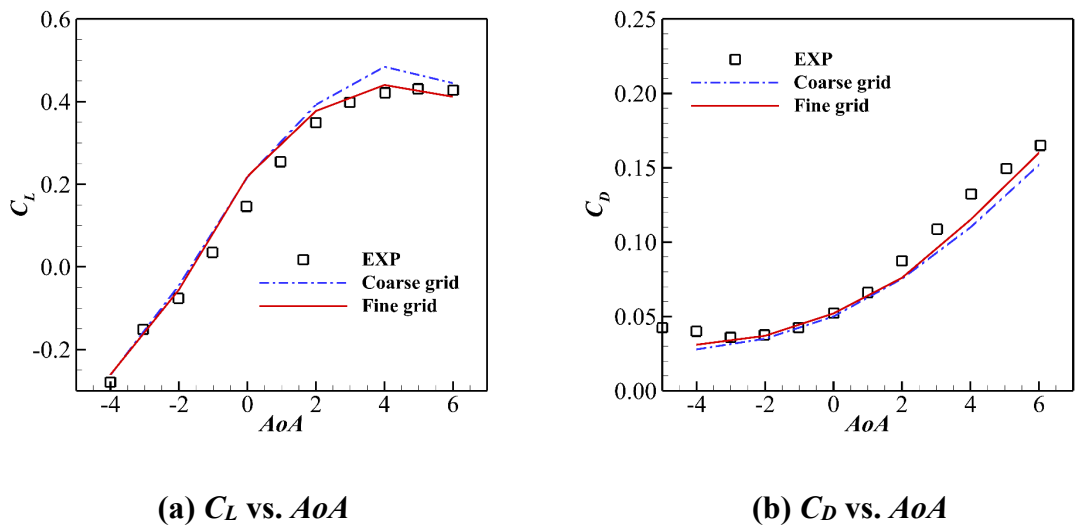

(a) $C_L$ vs. $AoA$      (b) $C_D$ vs. $AoA$

**Fig. 7. Aerodynamic coefficients prediction on different grid resolutions of the original finite-span wing geometry of Reghin et al.[20]**

Three configurations are then investigated (shown in Fig. 8): a clean infinite swept wing, an infinite swept wing with continuous ice, and an infinite swept wing with discontinuous ice. For the discontinuous-ice case, one ice segment together with its two adjacent gaps constitutes an ice-gap unit. The computational domain of the infinite discontinuous-ice configuration contains 25 such ice-gap units. Periodic boundary conditions are enforced in the spanwise direction to simulate the infinite-span swept wing. Structured grids are employed to better resolve the flow structures around the iced wings. The sectional and spanwise grid settings of the fine mesh used for the straight wing are adopted for the subsequent three infinite-span swept-wing configurations. Table 2 describes the grid details for the three infinite-span swept-wing configurations, using the same parameter definitions as in Table 1. The total grid numbers $N_{total}$ for the clean, continuous-ice, and discontinuous-ice wings are $84 \times 10^6$, $76 \times 10^6$ and $80 \times 10^6$, respectively. Figure 9 shows the grid for the infinite-span swept wing with discontinuous ice.

To characterize the discontinuity, the duty cycle concept introduced by Reghin et al.[20] is adopted. The duty cycle $\gamma_d$ is defined as follows:

$$\gamma_d = \frac{L_{gap}}{L_{ice} + L_{gap}} \qquad (14)$$

where $L_{gap}$ denotes width of the gap, while $L_{ice}$ represents the width of the ice. In this study, the duty cycle $\gamma_d$ is 52.8%.

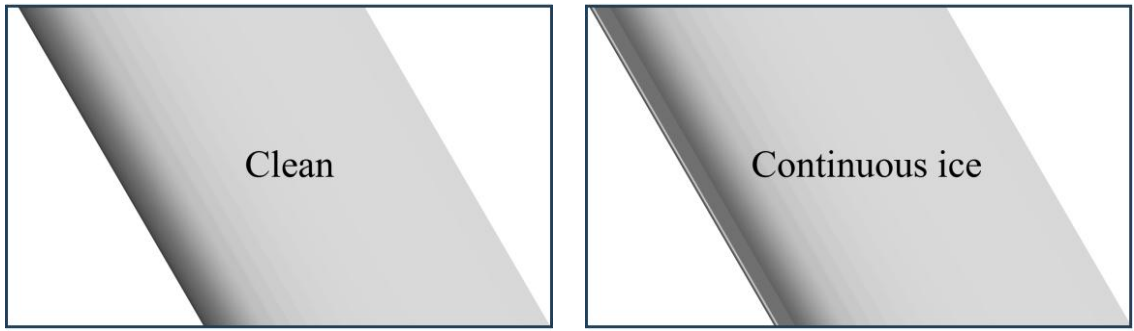

(a) clean infinite-span wing　　(b) Infinite-span wing with continuous ice

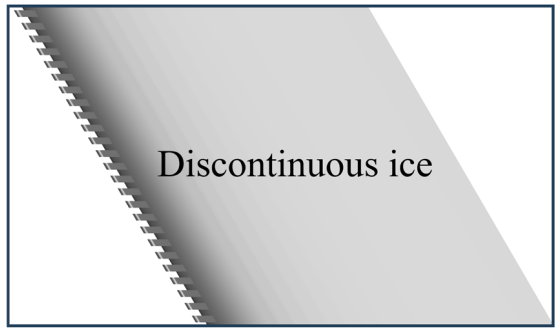

**(c) Infinite-span wing with discontinuous ice**

**Fig. 8. Geometries of the three infinite-span swept wings**

**Table 2: Grid details for three infinite-span swept wing configurations (Abbreviations: Con-ice = Continuous ice; Dis-ice = Discontinuous ice)**

| Grid | $N_x$ | $N_y$ | $N_z$ | $\Delta x_{ice}/c$ | $\Delta y_w/c$ | $\Delta z_w/c$ | $N_{total}$ |
|---|---|---|---|---|---|---|---|
| Clean | 482 | 253 | 875 | ------- | $3 \times 10^{-5}$ | 0.0013 | $84 \times 10^6$ |
| Con-ice | 482 | 253 | 875 | 0.0005 | $3 \times 10^{-5}$ | 0.0013 | $76 \times 10^6$ |
| Dis-ice | 482 | 253 | 875 | 0.0005 | $3 \times 10^{-5}$ | 0.0013 | $80 \times 10^6$ |

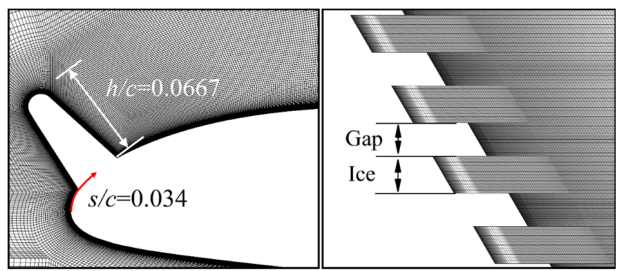

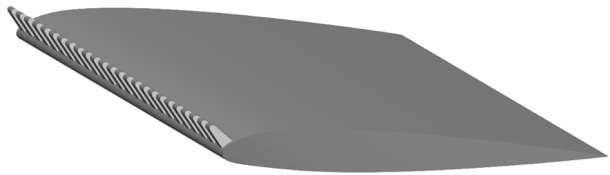

**Fig. 9. Grid for the infinite-span swept wing with discontinuous ice**

## 3.1 Aerodynamic force coefficients

The nondimensional physical time step is $\Delta t U_\infty/c = 0.0038$, with $U_\infty$ the freestream velocity and $c$ the chord length. The simulation is run for a total of 190 CTUs, and the last 76 CTUs are used for statistical analysis after the flow reaches a fully developed state. Figure 10 shows the temporal variations of the aerodynamic coefficients ($C_L$, $C_D$) for the continuous-ice and discontinuous-ice configurations at $AoA = 4°$ and $AoA = 8°$. The observed trend and periodic distribution of the force coefficients indicate that the simulation has converged. It can also be observed that the continuous-ice configuration induces much stronger unsteadiness in $C_L$ and $C_D$ compared with the discontinuous-ice case, particularly at $AoA = 4°$. This behavior suggests that continuous ice promotes large-scale flow separation, whereas discontinuous ice is associated with more localized, smaller-scale separation.

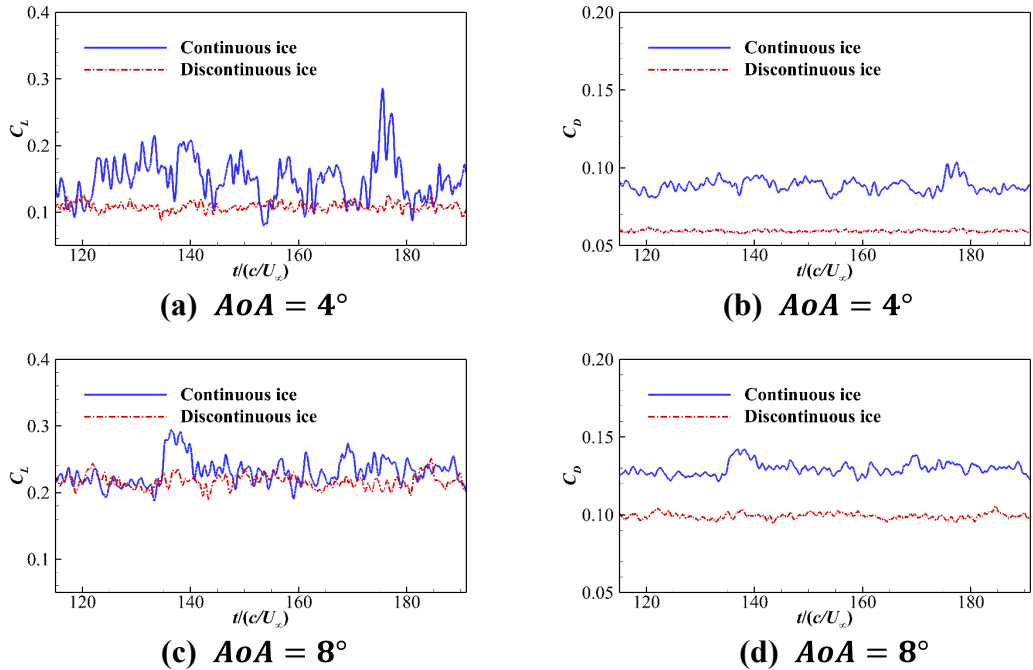

(a) $AoA = 4°$  (b) $AoA = 4°$

(c) $AoA = 8°$  (d) $AoA = 8°$

**Fig. 10. Fluctuations of aerodynamic coefficient ($C_L$, $C_D$) caused by continuous ice**

### and discontinuous at $AoA = 4°$ and $AoA = 8°$

To further examine the temporal characteristics of the lift and drag fluctuations ($C_L$ and $C_D$), power spectral density (PSD) analyses are performed for both the continuous-ice and discontinuous-ice swept wings, as shown in Figure 11. The horizontal axis denotes the nondimensional frequency, expressed as the Strouhal number $St = fc/U_\infty$, where $c$ is the chord length. At $AoA = 4°$, the discontinuous-ice case exhibits a characteristic frequency of $St$ = 22.6 in both $C_L$ and $C_D$, suggesting the presence of a periodic flow phenomenon induced by the discontinuous ice. As the $AoA$ increases to 8°, this characteristic frequency disappears. This is attributed to the formation of a large separation bubble at higher $AoA$, which dominates the flow field and weakens the unsteady effects associated with the discontinuous ice. For the continuous-ice case, no dominant frequency is observed.

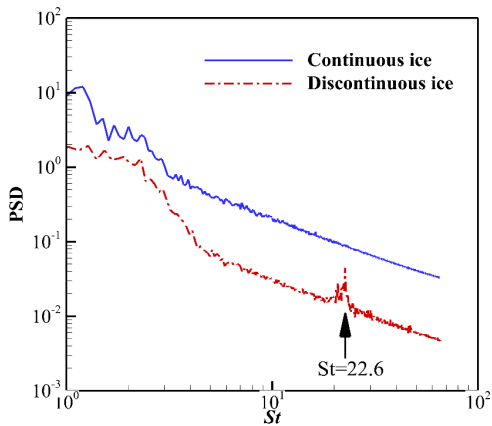 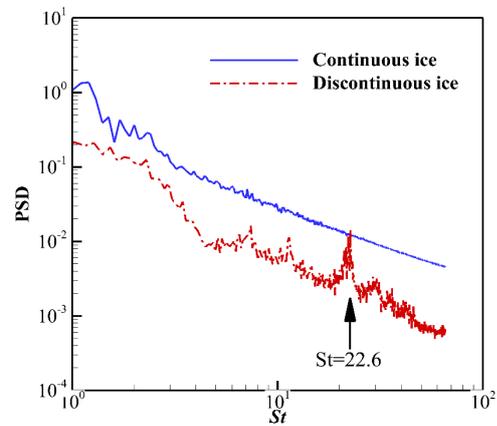

(a) $C_L$ at $AoA = 4°$     (b) $C_D$ at $AoA = 4°$

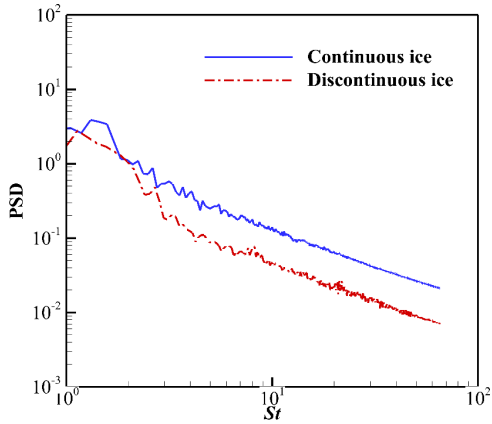 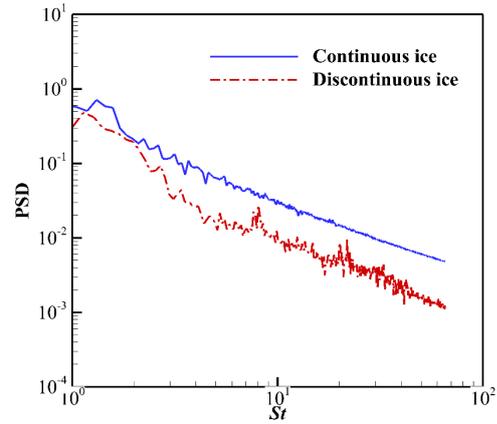

**(c)** $C_L$ at $AoA = 8°$        **(d)** $C_D$ at $AoA = 8°$

**Fig. 11. PSD for $C_L$, $C_D$ fluctuations of swept wings with continuous ice and discontinuous ice at $AoA = 4°$ and $AoA = 8°$**

Figure 12 shows the aerodynamic coefficients of swept wings with different leading-edge configurations. Figure 12(a) presents the lift coefficients for the clean leading edge, the continuous-ice leading edge, and the discontinuous-ice leading edge. The stall angles for these cases are 6°, 2°, and 0°, respectively. The relatively low lift coefficient of the clean wing is attributed to the infinite-span effect, which causes a larger flow separation compared with a finite-span wing, as well as the relatively low Reynolds number in this study. Compared with the continuous-ice case, the discontinuous ice causes stall to occur even earlier. Beyond an angle of attack of 12°, the differences among the three configurations become negligible because flow separation originates directly from the leading edge. An interesting observation is that, for the discontinuous-ice configuration, the lift coefficient does not exhibit a sharp drop after the stall angle. This is because the discontinuous ice disturbs the leading-edge flow even at very small *AoA*, resulting in turbulent flow throughout. Figure 12(b) shows that the drag coefficients follow the trend: clean leading edge < discontinuous-ice leading edge < continuous-ice leading edge. For the clean leading edge, the drag coefficient exhibits a sudden increase from $AoA = 6°$ to $AoA = 8°$, which is associated with stall onset.

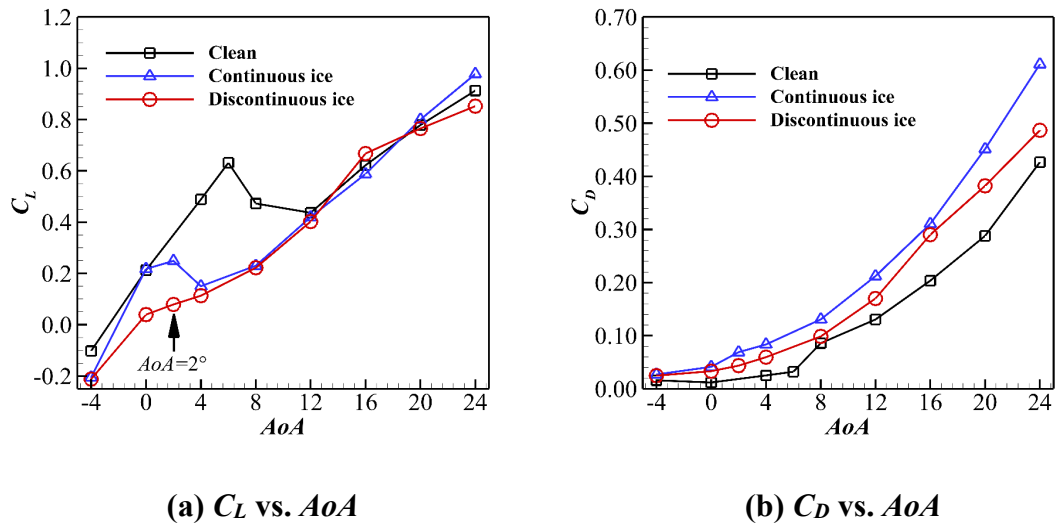

**(a)** $C_L$ **vs.** *AoA*          **(b)** $C_D$ **vs.** *AoA*

**Fig. 12. Aerodynamic coefficients on swept wings with different leading-edge configurations**

The lift coefficients of the continuous-ice and discontinuous-ice leading edges differ significantly at $AoA = 2°$, as shown in Figure 12(a). To clarify this difference, Figure 13 compares the pressure coefficient $C_p$ distributions between the continuous-ice and discontinuous-ice cases at $AoA = 2°$. For the continuous-ice case, a long suction plateau can be observed. In contrast, the discontinuous-ice case does not exhibit such a plateau.

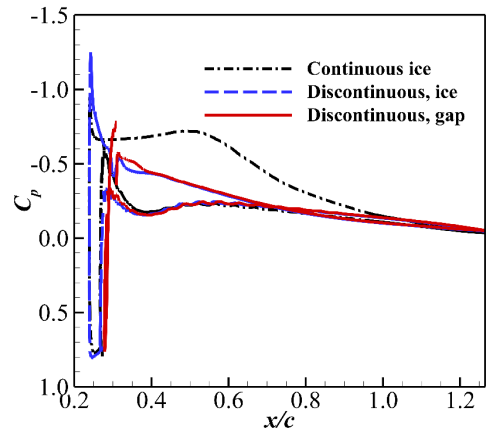

**Fig. 13. Comparison of pressure coefficient $C_p$ distributions between continuous-ice and discontinuous-ice case at $AoA = 2°$**

The variation in the pressure coefficient distribution arises from differences in flow behavior and separation patterns. Figure 14 shows the dimensionless velocity $U_x/U_\infty$ for the continuous-ice and discontinuous-ice wings at $AoA = 2°$. For the continuous-ice wing, a long separation bubble can be observed, which explains the extended suction plateau in Figure 13. In contrast, in the discontinuous-ice case, only a very small separation (behind the ice) is observed. The presence of discontinuous ice interferes with the formation of a large separation bubble, resulting in the absence of a distinct suction plateau in the pressure coefficient distribution.

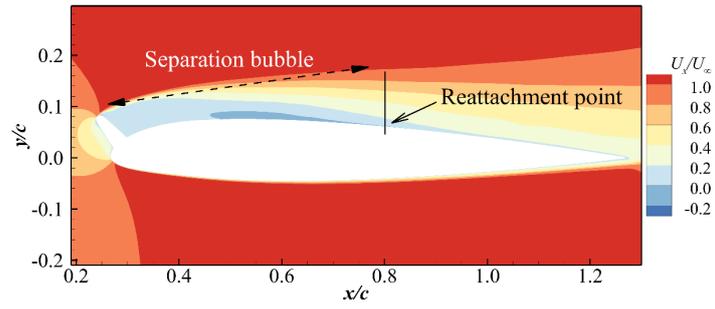

**(a) Continuous-ice wing**

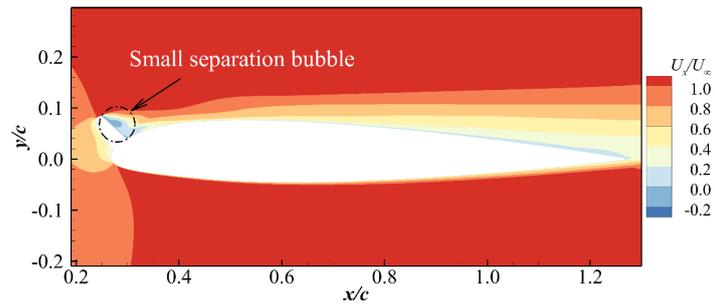

**(b) Discontinuous-ice wing**

**Fig. 14. The dimensionless velocity $U_x/U_\infty$ for continuous-ice and discontinuous-ice wings at $AoA = 2°$**

Figure 15 shows the dimensionless turbulence kinematic energy $k/U_\infty^2$ for the continuous-ice and discontinuous-ice wings at $AoA = 2°$. For the continuous-ice wing, the flow evolves sequentially from laminar to transitional and finally to turbulent states after the ice tip. Compared with the continuous-ice configuration, the discontinuous ice generates more complex flow structures, leading to strong turbulence at the leading edge. Consequently, the discontinuous-ice wing is less susceptible to flow separation.

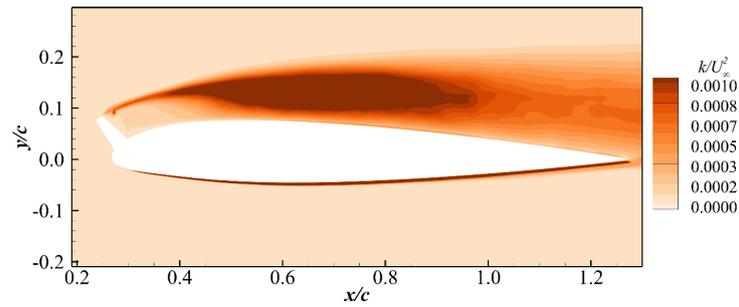

**(a) Continuous-ice wing**

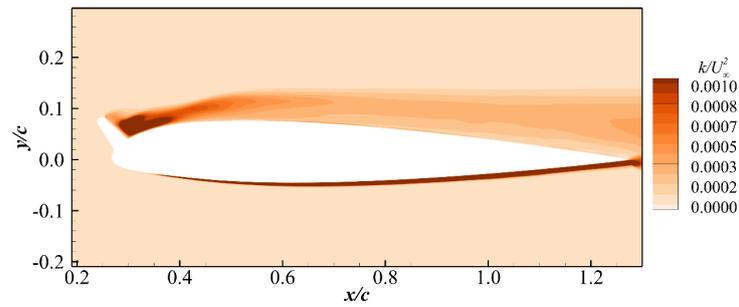

**(b) Discontinuous-ice wing, ice section**

**Fig. 15. The dimensionless turbulence kinematic energy $k/U_\infty^2$ for discontinuous ice at**

$$AoA = 2°$$

## 3.2 Instantaneous field and vortex structures

Figure 16 shows the evolution of vortex structures over swept wings with different leading-edge configurations at $AoA = 2°, 4°$ and $8°$. In the case of the continuous-ice wing, Kelvin–Helmholtz (K–H) instability and transition occur within the separating shear layer immediately downstream of the ice tip. Further downstream, the hairpin vortices become densely packed and disordered. The continuous ice induces a leading-edge separation that originates from the ice tip. In contrast, the discontinuous-ice wing shows a marked absence of typical two-dimensional K–H instability and transition. The gap jets between ice segments introduce complex flow disturbances that rapidly disrupt the formation of 2D coherent vortical structures and influence the transition process. A more detailed examination of the vortex dynamics in the vicinity of the discontinuous ice will be presented later. As the angle of attack increases, noticeable differences emerge between the continuous-ice wing and the discontinuous-ice wing. For the continuous ice, the separation bubble remains on the upper surface and continues to expand in size. For the discontinuous ice, the separation bubble is still located near the trailing region but grows larger, extending over a greater portion of the upper surface toward the leading edge. For $AoA = 2°, 4°$ and $8°$, a similar phenomenon is observed: the continuous-ice configuration produces larger hairpin vortices than the discontinuous-ice swept wing. This can be attributed to two main factors: (1) the higher velocity over the upper surface, which corresponds to a larger local Reynolds number, and (2) the dominant eddies being located farther from the wall in the continuous-ice case.

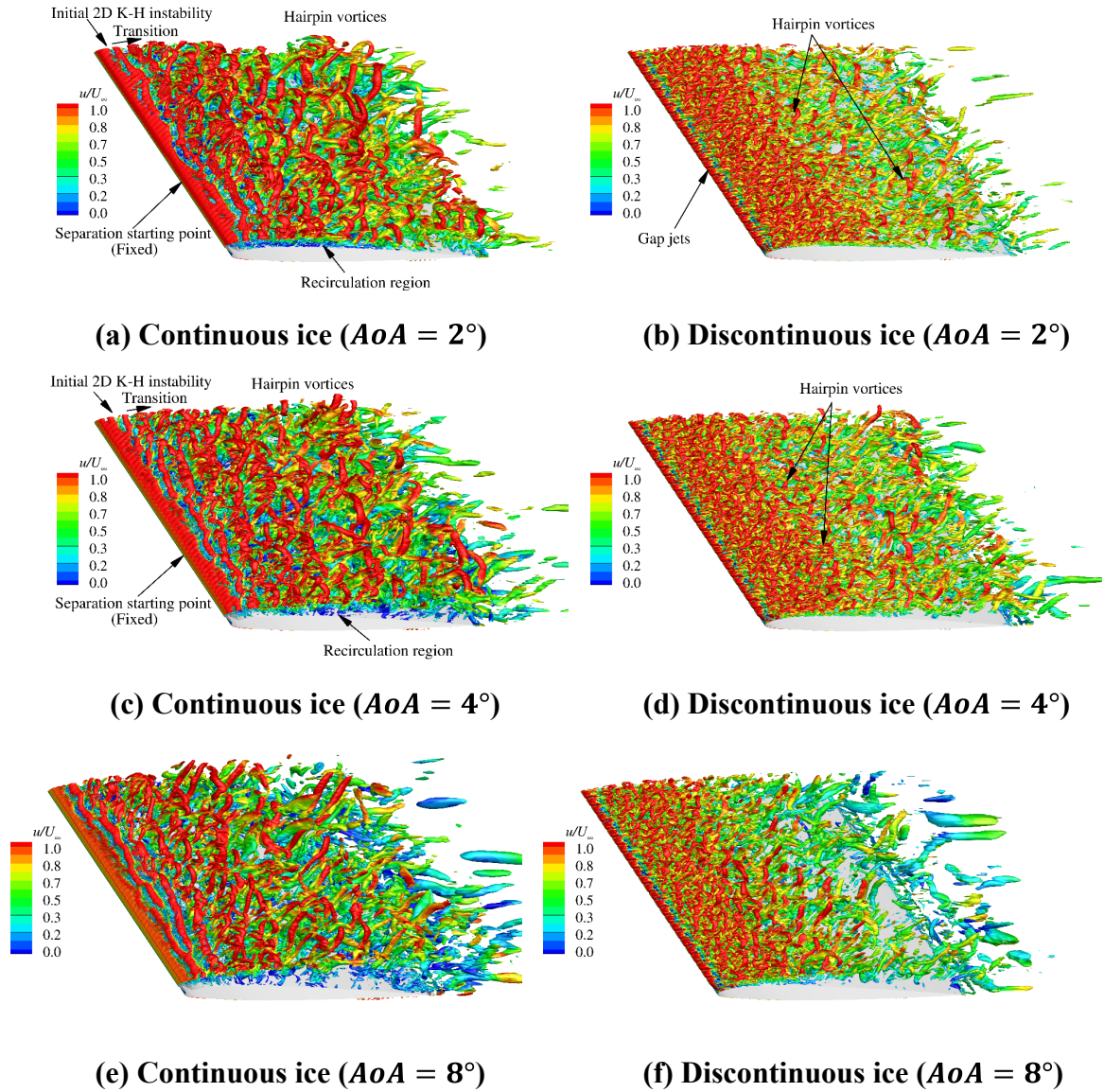

(a) Continuous ice ($AoA = 2°$)  (b) Discontinuous ice ($AoA = 2°$)

(c) Continuous ice ($AoA = 4°$)  (d) Discontinuous ice ($AoA = 4°$)

(e) Continuous ice ($AoA = 8°$)  (f) Discontinuous ice ($AoA = 8°$)

**Fig. 16. Evolution of vortex structures ($Q(c/U_\infty)^2 = 100$) over swept wings with different leading-edge configurations at $AoA = 2°, 4°,$ and $8°$**

Figure 17 presents the generation and evolution of vortical structures induced by continuous ice at $AoA = 4°$. The vortical structures are visualized using the nondimensional Q-criterion $Q(c/U_\infty) = 30$, together with the nondimensional streamwise vorticity $\omega_x(c/U_\infty)$. In Figure 17(a), the presence of discontinuous ice at the leading edge generates multiple pairs of counter-rotating vortices, which quickly promote the transition of the leading-edge boundary layer from laminar to turbulent. Figure 17(b) shows the vortex structures generated by individual ice elements. Initially, two distinct counter-rotating vortices form. As they convect downstream, their mutual interactions lead to rapid breakdown, resulting in a cluster of disorganized vortical structures.

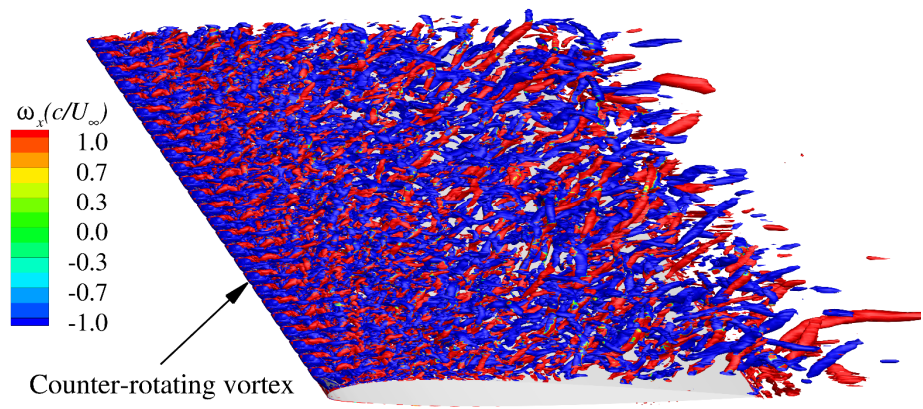

**(a) Overall vortices evolution**

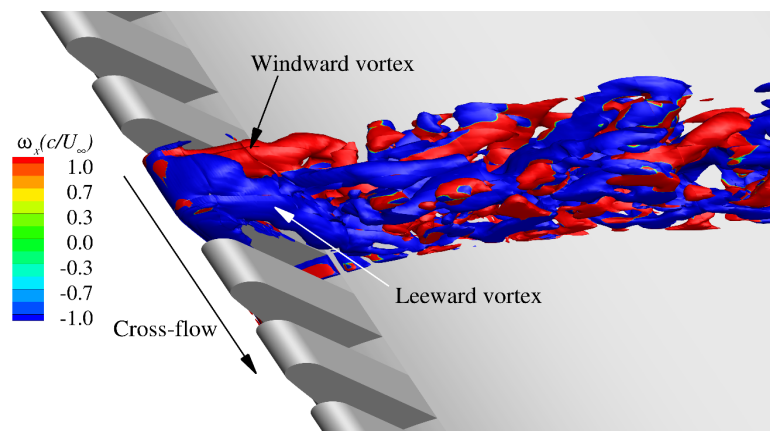

**(b) Closeup of leading vortices**

**Fig. 17. generation and evolution of vortical structures ($Q(c/U_\infty)^2 = 100$) induced**

<h2 style="text-align:center">by continuous ice at $AoA = 4°$</h2>

Continuous and discontinuous ice shapes lead to different characteristics in the separating shear layer and leading-edge flow. To better illustrate these differences, the triple decomposition of the velocity gradient is employed in the following analysis. This approach helps identify key variations in the separating shear layer, where pure shearing dominates the velocity gradient. Traditionally, the velocity gradient tensor $G_{ij}$ (VGT) is partitioned into a symmetric strain-rate tensor and an antisymmetric vorticity tensor. A more refined framework, known as triple decomposition, partitions the VGT into three physically distinct components: normal straining, pure shearing, and rigid-body rotation. Arun et al.[35] applied triple decomposition to isotropic turbulence, yielding deeper insights into its flow structures. In the present study, we employ this approach to iced wings, where it reveals additional flow features that are not captured by the conventional decomposition. The VGT can be expressed in its principal reference frame. In this frame, the VGT is quasi-triangular, and it can be decomposed as:

$$G_{ij} = G_{ij}^N + G_{ij}^R + G_{ij}^S = \begin{bmatrix} \varepsilon_1 & 0 & 0 \\ 0 & \varepsilon_2 & 0 \\ 0 & 0 & \varepsilon_3 \end{bmatrix} + \begin{bmatrix} 0 & 0 & 0 \\ 0 & 0 & \varphi_1 \\ 0 & -\varphi_1 & 0 \end{bmatrix} + \begin{bmatrix} 0 & \gamma_3 & \gamma_2 \\ 0 & 0 & \gamma_1 \\ 0 & 0 & 0 \end{bmatrix} \qquad (15)$$

where $G_{ij}^N$, $G_{ij}^R$ and $G_{ij}^S$ denote the normal straining, rigid rotation and pure shearing tensors, respectively. These tensors can be determined and transformed to the original coordinate system using the ordered real Schur decomposition of $G_{ij}$[36].

Correspondingly, the strength of the velocity gradients can be expressed as

$$G_{ij}G_{ij} = G_{ij}^{N}G_{ij}^{N} + G_{ij}^{R}G_{ij}^{R} + G_{ij}^{S}G_{ij}^{S} + G_{ij}^{R}G_{ij}^{S} \tag{16}$$

The first three terms represent the strengths of the constituents in Eqn. (15), and the last term represents the interaction between shearing and rigid rotation. Furthermore, the velocity gradient partitioning is defined in terms of the relative contributions of these constituents to $G_{ij}G_{ij}$. The relative contribution of normal straining is $gg_{ns} = G_{ij}^{N}G_{ij}^{N}/G_{ij}G_{ij}$, the relative contribution of rigid rotation is $gg_{rr} = G_{ij}^{R}G_{ij}^{R}/G_{ij}G_{ij}$, the relative contribution of pure shearing is $gg_{ps} = G_{ij}^{S}G_{ij}^{S}/G_{ij}G_{ij}$, and the relative contribution of the interaction between shearing and rigid is $gg_{rs} = G_{ij}^{R}G_{ij}^{S}/G_{ij}G_{ij}$.

Triple decomposition is employed to reveal the separating shear layer and provide more detailed insight into the flow structures. Figure 18 presents the triple decomposition of the velocity gradient for the continuous-ice and discontinuous-ice swept wings at $AoA = 4°$. This decomposition clearly illustrates the spatial distribution of pure shearing, rigid-body rotation, and normal straining in the downstream region of the ice shape. A pronounced difference is observed in the characteristics of the separating shear layer. The continuous-ice wing exhibits a stronger and more coherent separating shear layer with a smooth, well-defined curvature, within which pure shearing is dominant. In contrast, for the discontinuous-ice case, the presence of the gap jets disrupts the separation bubble, leading to a more irregular and distorted separating shear layer. Figure 18 also provides detailed insight into other flow structures. The rigid-rotation structures are surrounded by pure-shearing layers, which agrees with the description of turbulent structures reported by Arun et al.[36]. In addition, normal straining persists in the flow downstream of the ice shape and dominates the velocity gradient in the outer mainstream region away from the wall.

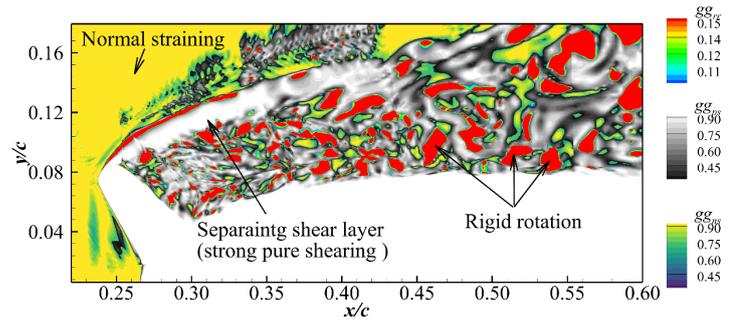

**(a) Continuous-ice swept wing**

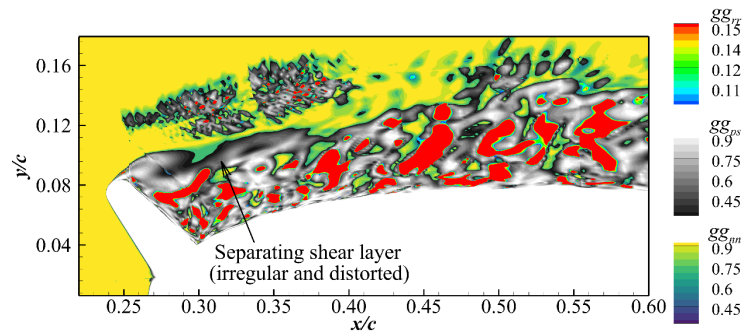

**(b) Discontinuous-ice swept wing (ice section)**

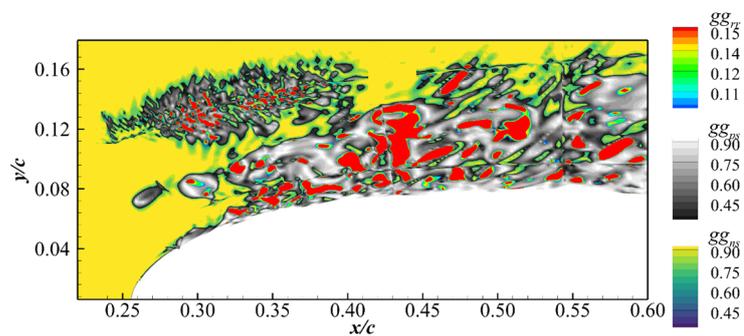

**(c) Discontinuous-ice swept wing (gap section)**

**Fig. 18. Velocity gradient partitioning in for continuous-ice and discontinuous-ice swept**

**wings at $AoA = 4°$**

Fig. 19 presents the spanwise vorticity distributions of swept wings with three different leading-edge configurations at $AoA = 4°$ (left column) and $8°$ (right column). The spanwise vorticity contours reveal distinct flow separation characteristics between the clean and iced airfoils. For the clean configuration, the flow remains largely attached at $AoA = 4°$, with only weak vorticity observed near the trailing edge (Fig. 19 (a)). At $AoA = 8°$, however, stall characteristics can be observed (Fig. 19(b)). In contrast, the continuous-ice case (Figs. 19(c) and 19(d)) exhibits a separation bubble even at small angles of attack, originating from the ice location. Downstream of the ice tip, the Kelvin–Helmholtz instability is clearly visible, and the separating shear layer becomes well defined, forming a typical fixed-separation-point bubble structure. Increasing the angle of attack causes the vortex shedding to shift away from the surface, reducing its interaction with the wall. For the discontinuous-ice case (Figs. 19(e) and 19(f)), the vorticity field lacks organized or coherent structures, displaying a more fragmented and irregular pattern. This indicates that the gap jets prevent the formation of a large, coherent separation bubble near the leading edge by breaking up the shear layer into smaller structures. As a result, the discontinuous ice induces multiple characteristic frequencies and exhibits stronger three-dimensional flow effects compared with the continuous-ice case.

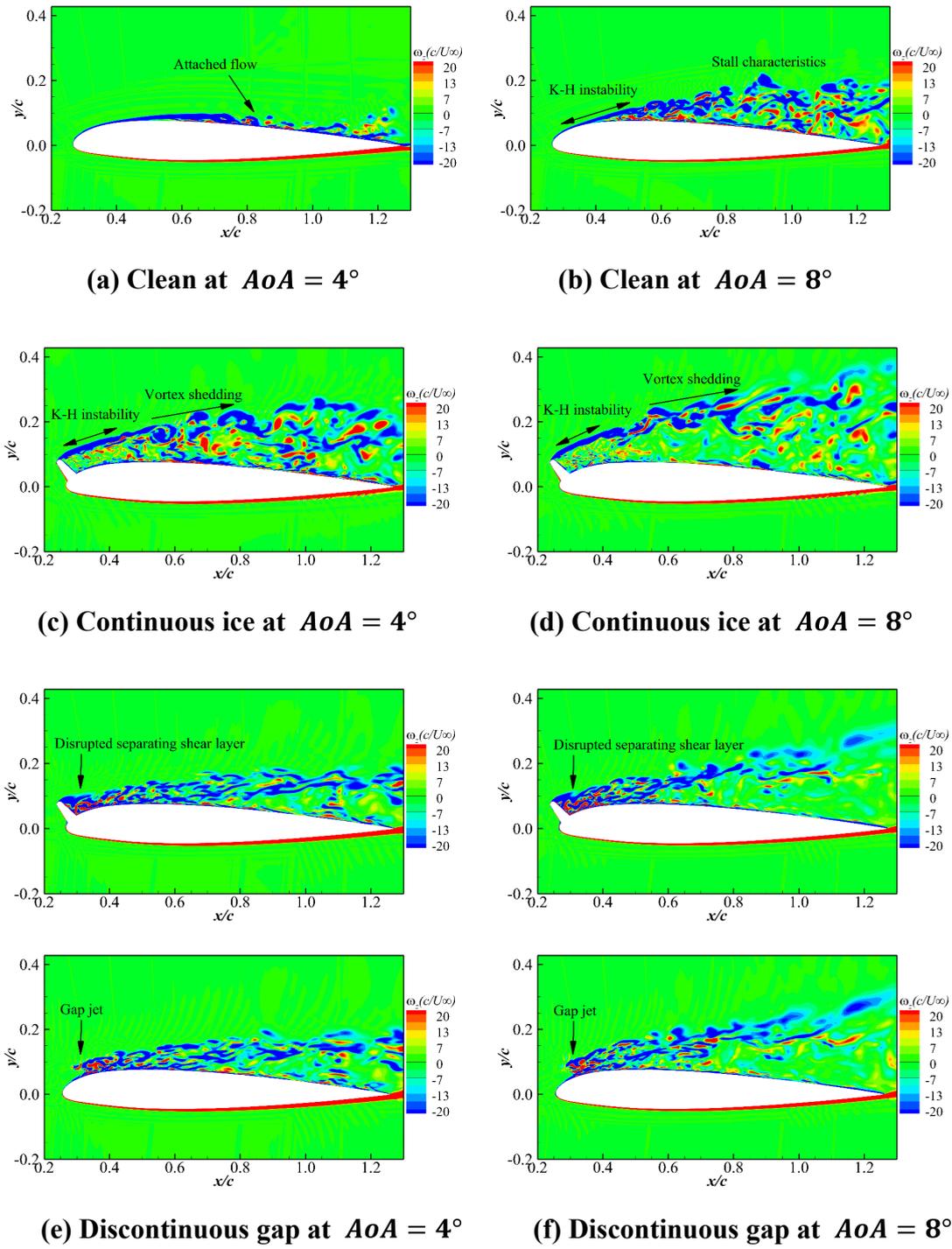

**(a) Clean at $AoA = 4°$**
**(b) Clean at $AoA = 8°$**
**(c) Continuous ice at $AoA = 4°$**
**(d) Continuous ice at $AoA = 8°$**
**(e) Discontinuous gap at $AoA = 4°$**
**(f) Discontinuous gap at $AoA = 8°$**

**Fig. 19. Spanwise vorticity distributions of swept wings with three different leading-edge configurations at $AoA = 4°$ and $AoA = 8°$**

## 3.3 Frequencies and flow modes analysis

Owing to the presence of two distinct separation regions—a small one immediately downstream of the ice accretion and a larger one extending over the upper surface—7 probe points were strategically placed to capture the essential flow features. Fig. 20 shows the probe-point locations for the discontinuous-ice case at $AoA = 4°$. These points were chosen to analyze the characteristic flow dynamics and dominant frequencies.

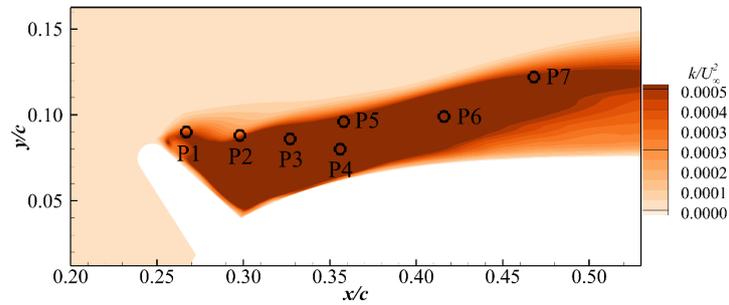

(a) Points at ice section

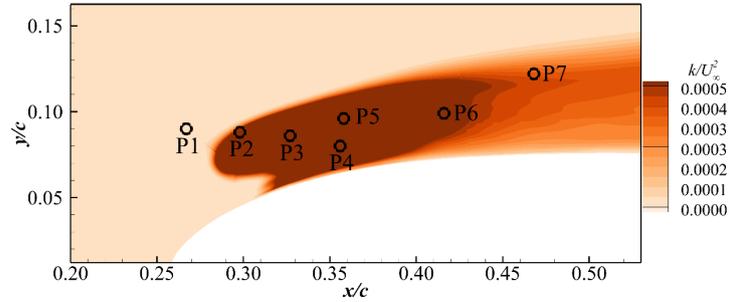

(b) Points at gap section

**Fig. 20. Probe locations for discontinuous ice at $AoA = 4°$**

Fig. 21 presents the PSD of the probe points at both the ice section and the gap section. Three characteristic frequencies, $St_1 = 11.3$, $St_2 = 22.6$, and $St_3 = 33.9$, are identified, exhibiting a harmonic relationship. The emergence of multiple frequencies is attributed to structural breakup, which generates progressively smaller vortical structures corresponding to higher harmonics. These high-frequency structures dissipate rapidly as they convect downstream. Specifically, at probe locations P6 and P7, the third harmonic ($St_3 = 33.9$) nearly vanishes, indicating that the small-scale structures associated with this frequency have been largely dissipated.

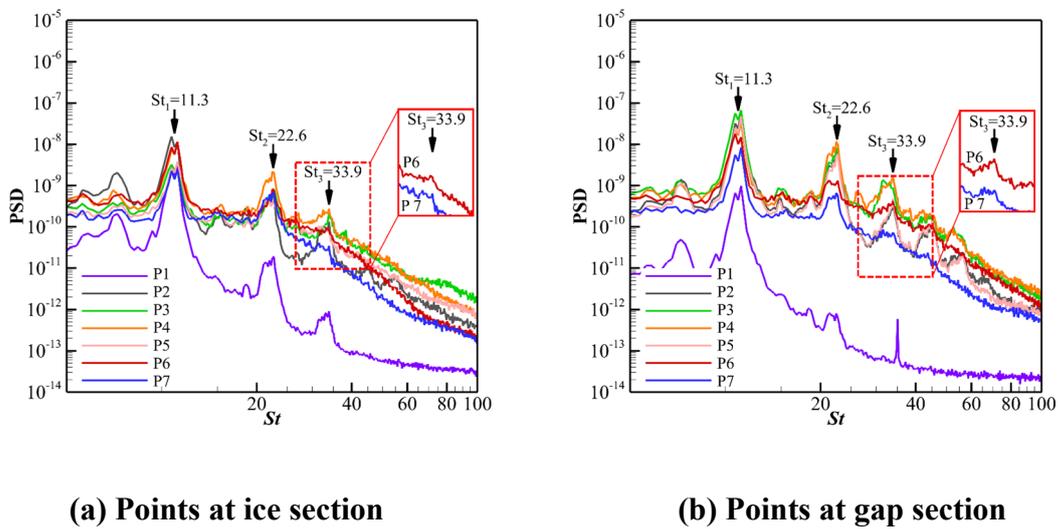

(a) Points at ice section       (b) Points at gap section

**Fig. 21. Power spectral density (PSD) for the probe points at both the ice section and gap section of the discontinuous-ice wing at $AoA = 4°$**

Fig. 22 illustrates the instantaneous evolution of vortical structures ($Q(c/U_\infty)^2 = 100$, contoured by $\omega_x c/U_\infty$) over the discontinuous ice at $AoA = 4°$. The characteristic time scale is $\Delta T_1 = 1/f_1 = c/(U_\infty * St_1)$. Panels (a)–(d) correspond to $t$, $t + 1/3\Delta T_1$, $t + 2/3\Delta T_1$ and $t + \Delta T_1$, respectively. A pair of counter-rotating vortices is shed from the leading discontinuous ice. The windward vortex (shown in red) undergoes a complete evolution process, including its generation, downstream convection, breakup, and eventual dissipation. During the breakup stage, it divides into two smaller vortices corresponding to a higher characteristic frequency ($St_2 = 22.6$). On the leeward side, the vortex (shown in blue) is sequentially generated, stretched by the local shear, and gradually elongated along the streamwise direction before breaking up into smaller-scale structures.

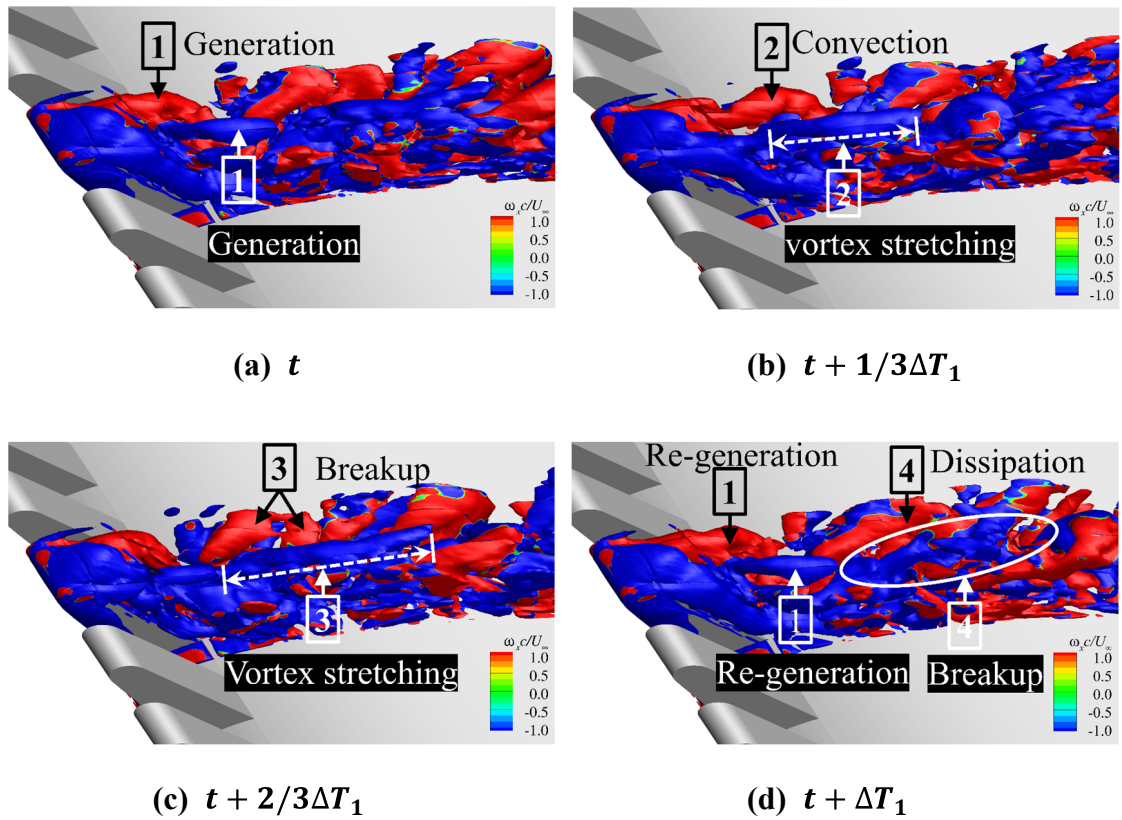

**Fig. 22**. **Instantaneous vortex evolution process of the discontinuous ice at $AoA = 4°$**

($Q(c/U_\infty)^2 = 100$, **contoured by** $\omega_x c/U_\infty$)

The observed vortex shedding closely resembles the wake dynamics of a circular cylinder. By analogy, if the characteristic length in $St_1$ is replaced by an equivalent diameter of the ice feature rather than the airfoil chord, the scaling becomes comparable to that of a cylinder wake.

In the present study, ice dimensions are $h/c = 0.0667$ and $w/c = 1.96\%$. Taking the root width as $l/c \approx 0.0334$ and defining an equivalent diameter as $d_e = \sqrt{cw} = 0.0256$, the Strouhal number based on this length scale is $St_1^* = fd_e/U_\infty = 0.58$. Brown[37] reported a nearly universal Strouhal number of about 0.176 (ranging from 0.164 to 0.186) for circular cylinders at Reynolds numbers between 55 and $1.4 \times 10^5$. The cylinder-wake vortex-shedding frequency serves only as a reference. The vortex shedding of discontinuous ice has a characteristic Strouhal number of 0.58, which is larger but of the same order of magnitude as that for a canonical cylinder flow.

Proper Orthogonal Decomposition (POD), also known as Karhunen–Loeve decomposition in mathematics, provides a method to decompose a dataset into a set of orthogonal modes with corresponding temporal coefficients and eigenvalues. POD facilitates the extraction of high-resolution snapshots from instantaneous flow fields for further analysis. In this study, 1405 sampling points were selected for eigenvalue analysis of the iced swept wing. The convergence of the dataset was previously validated in Refs[38][39].

In POD analysis, velocity and pressure are commonly used physical quantities. Preliminary investigations indicate that velocity-based POD fails to effectively capture shear-layer separation and vortex evolution. This limitation arises from the strong three-dimensionality and spanwise development of the flow over the iced swept wing, which cannot be fully represented in the two-dimensional $x-y$ plane[40]. Therefore, pressure is selected for the subsequent POD analysis, as it more clearly reveals the evolution of vortex structures, as discussed in the following sections.

Fig. 23 and Table 3 show the cumulative energy contribution and energy fraction of the POD modes, providing a quantitative basis for selecting the dominant modes that capture the majority of energy in the fluctuating flow field. The first 20 modes account for about 63% of the total energy in the ice section, whereas they capture approximately 80% in the gap section. Zhou et al.[25] reported that in continuous-ice swept wings, the first 20 modes represent nearly 90% of the total flow energy. This broader distribution of energetic modes in the present discontinuous-ice case indicates a more complex flow field, where the flow energy is not concentrated in a few dominant modes but dispersed across many. Such dispersion implies weaker large-scale coherent structures and the coexistence of multiple interacting flow patterns. Consequently, the modal efficiency is lower, as more modes are required to represent the same portion of total energy.

**Table 3: Cumulative model contribution of POD modes.**

| Slice | Mode1 | First 5 modes | First 10 modes | First 20 modes | First 50 modes |
|---|---|---|---|---|---|
| Ice section | 12% | 37% | 49% | 63% | 81% |
| Gap section | 29% | 62% | 71% | 80% | 90% |

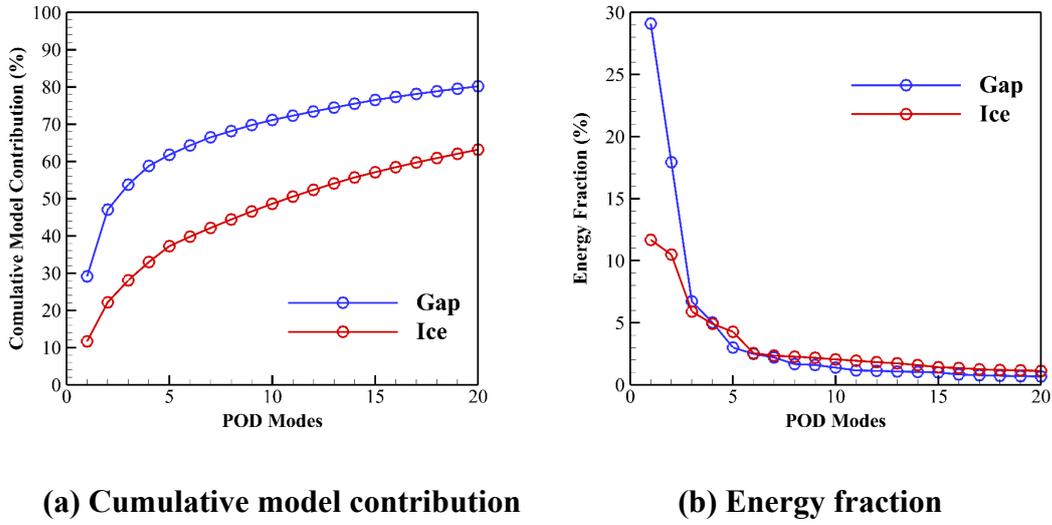

**(a) Cumulative model contribution**   **(b) Energy fraction**

**Fig. 23. Cumulative model contribution and energy fraction of POD modes for discontinuous ice at $AoA = 4°$**

Fig. 24 presents the first four POD mode coefficient fields for the discontinuous-ice configuration at $AoA = 4°$. In this analysis, the pressure coefficient is used instead of the velocity component. Fig. 24(a) presents the first four POD spatial modes at the ice section. The first mode pair (Modes 1 and 2) represents low-frequency, large-scale coherent structures that exhibit regular patterns along the upper surface. These two modes are spatially shifted by approximately one-quarter wavelength, suggesting a quasi-periodic motion of the flow structures. This mode pair captures the dominant periodic motion associated with the frequency $St_1 = 11.3$. Modes 3 and 4 show large structures dominated by spanwise flow downstream of $x/c = 0.10$. Fig. 24(b) presents the first four POD spatial modes at the gap section. The first mode pair (Modes 1 and 2) shows the dominant periodic motion with a characteristic frequency of $St_1 = 11.3$, while the second pair (Modes 3 and 4) corresponds to a higher-frequency periodic motion at $St_2 = 22.6$. The first pair primarily captures large-scale, low-frequency flow structures, whereas the second pair captures smaller-scale, higher-frequency flow features. Modes 3 and 4 are dominated by the gap jet rather than the spanwise flow, leading to more regular structural patterns. In Mode 3, flow structures also undergo breakup, generating smaller-scale, higher-frequency components ( $St_2 = 22.6$ ); further downstream, these structures re-merge, forming more regular patterns.

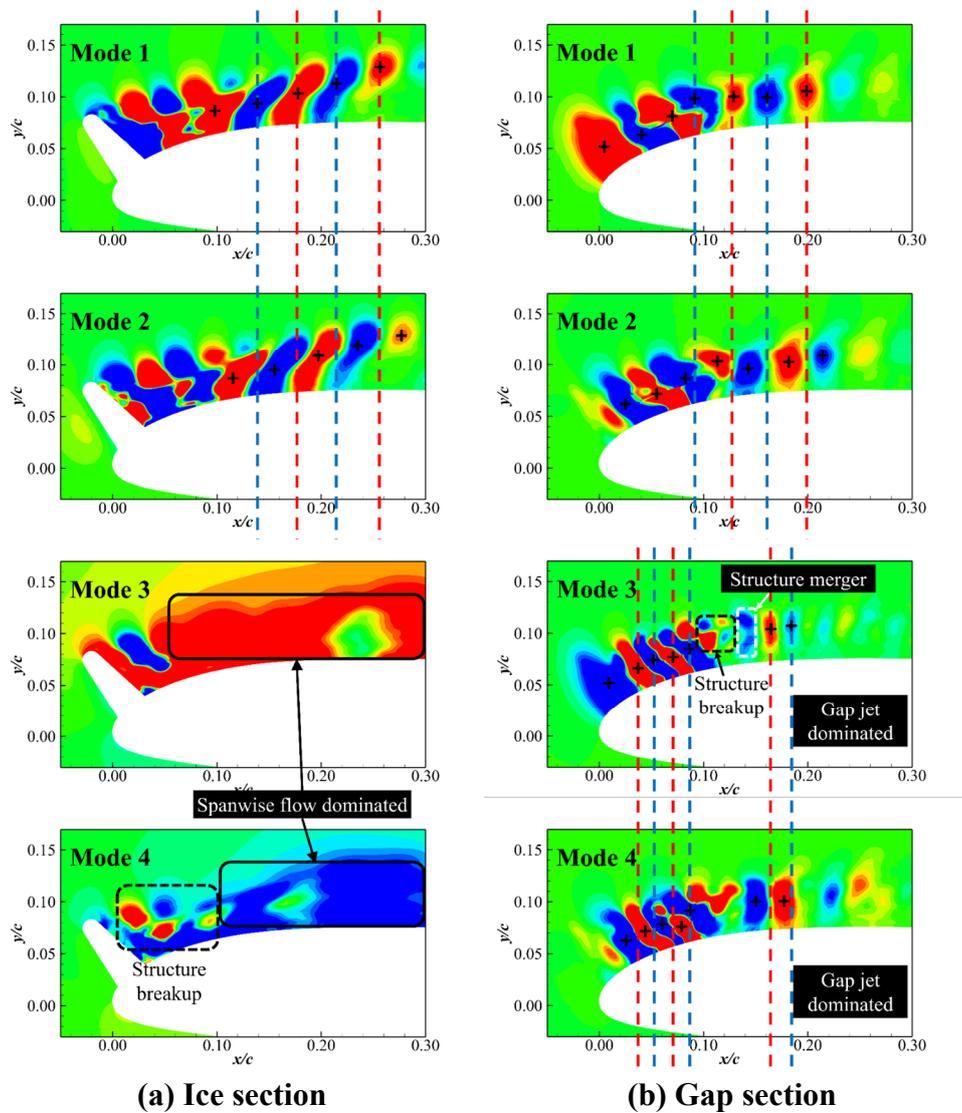

**(a) Ice section**        **(b) Gap section**
**Fig. 24. The first four mode POD coefficient fields for discontinuous ice at $AoA = 4°$**

Fig. 25 shows the schematic of the flow structures around the discontinuous ice. This complex flow can be conceptually simplified as a combination of two canonical patterns: a separating shear layer and a Kármán vortex street. The separating shear layer originates from the ice tip, while the Kármán vortex street is shed from both sides of the ice. This simplified flow model provides a qualitative explanation for the frequency characteristics observed at both the ice section and the gap section. The Kármán vortex street exhibits three frequencies ($St_2 = 11.3$, $St_2 = 22.6$, $St_2 = 33.9$), reflecting its periodic shedding behavior. In contrast, the separating shear layer near the ice section differs markedly from that in the continuous-ice case. The strong interference from the gap jets disturbs the shear-layer development, leading to a highly irregular and distorted structure.

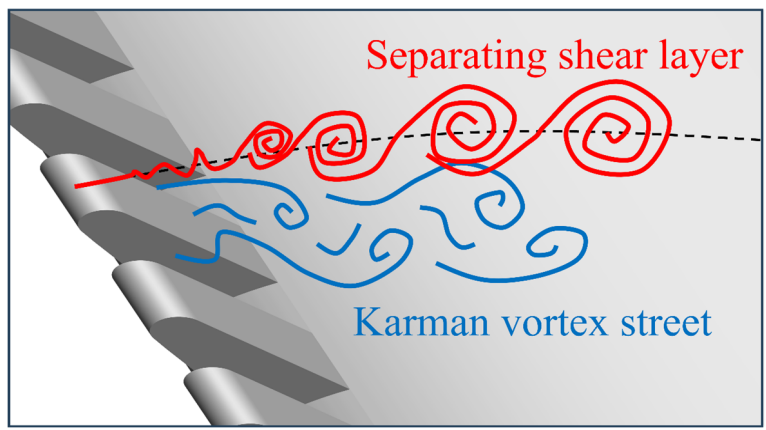

**Fig. 25. The schematic of the flow structures around the discontinuous ice**

Fig. 26 shows the PSD of the lift-force coefficient and the first four temporal coefficients decomposed by POD for the ice section at $AoA = 4°$. The lift force coefficient exhibits a characteristic frequency of $St_2 = 22.6$, as mentioned earlier. As shown in Fig. 26(a), two characteristic frequencies, 11.3 and 22.6, appear at the ice section, while Fig. 26(b) reveals three distinct frequencies—11.3, 22.6, and 33.9—at the gap section. Notably, although both the ice section and the gap section have the frequency $St_2 = 22.6$, the gap section shows higher PSD magnitudes than the ice section. The $St_2 = 22.6$ at the ice section is likely induced by flow dynamics originating from the gap region. This finding suggests that the gap region predominantly contributes to the characteristic frequency and to the fluctuation of the lift coefficient.

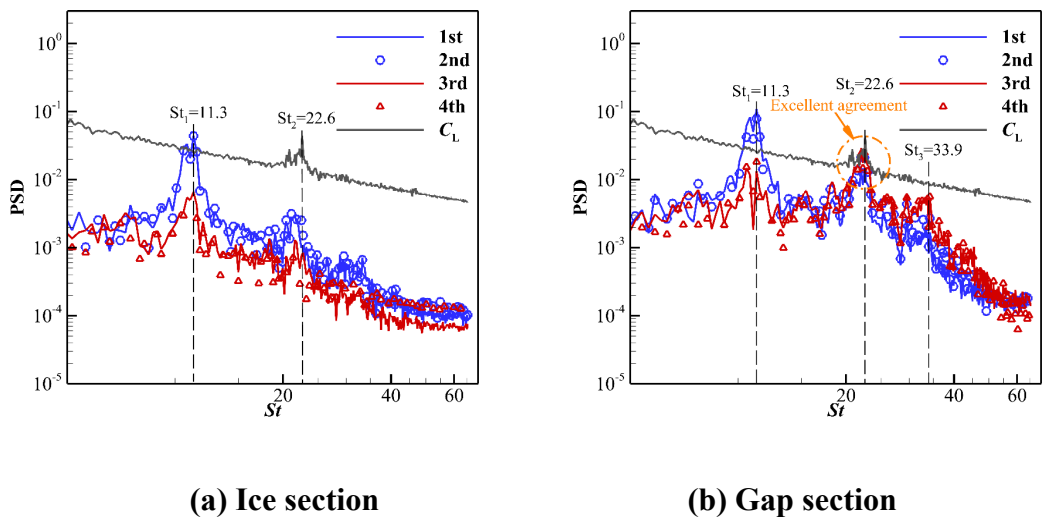

(a) Ice section  (b) Gap section

**Fig. 26. PSD of the lift force coefficient and the first four temporal coefficients decomposed with POD for discontinuous-ice swept wing at $AoA = 4°$**

## 4. Conclusions

This study investigates the effects of discontinuous ice through the AMD-IDDES method. Comparisons are made among a clean swept wing, a continuous-ice swept wing, and a discontinuous-ice swept wing. The analysis focuses on three aspects: aerodynamic force characteristics, leading-edge flow structures, and flow frequencies. Several conclusions are drawn from this work:

(1) Discontinuous ice has a more detrimental effect on aerodynamic performance than continuous ice. Compared with the clean wing, continuous ice leads to an earlier stall angle of attack, and discontinuous ice advances the stall angle even further. The underlying mechanisms differ: continuous ice induces a large separation bubble on the upper surface, where the associated low-pressure region helps to partially sustain lift. By contrast, the gap jets generated by discontinuous ice suppress the formation of such a bubble, resulting in a more pronounced loss of lift. Nevertheless, while discontinuous ice causes a greater reduction in lift, its penalty on drag is less severe than that imposed by continuous ice.

(2) The flow over the discontinuous-ice wing exhibits two primary features: a separating shear layer and Kármán vortex shedding. However, the separating shear layer becomes highly irregular due to the interference of gap jets between adjacent ice sections. In addition, the discontinuous ice generates counter-rotating vortex pairs that convect downstream, gradually disintegrate, and induce higher-frequency fluctuations in the wake. Spectral analysis reveals three characteristic chord-based Strouhal numbers—11.3, 22.6, and 33.9. The lowest-frequency component ($St$ =11.3) corresponds to the shedding of vortex pairs; when nondimensionalized by the ice width, it gives $St$ = 0.58, which exceeds that of a canonical cylinder wake, likely due to interactions between neighboring counter-rotating vortex pairs.

(3) The discontinuous-ice configuration exhibits a distinct frequency signature ($St$ = 22.6) in

the fluctuations of both lift and drag coefficients. In particular, the lift and drag coefficients oscillate at a frequency twice that of the counter-rotating vortex pairs, primarily due to the gap jets. In contrast, the continuous-ice configuration shows no such characteristic frequency in either lift or drag fluctuations.

This work offers detailed results on aerodynamic characteristics, vortex structures, and characteristic frequencies associated with discontinuous ice. However, the present work focuses on infinite-span swept wings to eliminate the influence of wing-root and wing-tip effects. Future studies should extend the investigation to finite wings to capture additional effects.

## Acknowledgments

This work was supported by the National Natural Science Foundation of China (grant numbers 12372288, U23A2069 and 12388101); and the National Key Research and Development Program of China (2024YFB4205601) and other national research projects.

## Author declarations

### Conflicts of Interest

The authors have no conflicts to disclose.

## Data availability

The data that support the findings of this study are available from the corresponding author upon reasonable request.